\documentclass[pdflatex,sn-mathphys-num]{sn-jnl}

\usepackage{amsmath, amssymb, amsfonts, amsthm, siunitx}

\usepackage{graphicx}
\usepackage{multirow}
\usepackage{booktabs}
\usepackage{array}
\usepackage{colortbl}
\usepackage{tabularx}
\usepackage{subcaption}
\usepackage[most]{tcolorbox}

\usepackage{algorithm}
\usepackage{algorithmicx}
\usepackage{algpseudocode}

\usepackage{xcolor}
\usepackage{textcomp}
\usepackage{url}
\usepackage{verbatim}
\usepackage{stfloats}

\theoremstyle{thmstyleone}%
\theoremstyle{thmstyletwo}%

\theoremstyle{thmstylethree}%

\begin{document}
\title{Simulation of Language Evolution under Regulated Social Media Platforms: A Synergistic Approach of Large Language Models and Genetic Algorithms}

\author[1]{\fnm{Jinyu} \sur{Cai}}

\author[2]{\fnm{Mingyue} \sur{Zhang}}

\author[3]{\fnm{Yusei} \sur{Ishimizu}}

\author[4]{\fnm{Munan} \sur{Li}}

\author*[1]{\fnm{Jialong} \sur{Li}}

\author[3]{\fnm{Kenji} \sur{Tei}}

\begingroup
\renewcommand\thefootnote{}\footnotetext{Author contact details and profiles are listed in the “Authors and Affiliations” section at the end of this article.}%
\addtocounter{footnote}{-1}%
\endgroup






\abstract{
Social media platforms frequently impose restrictive policies to moderate user content, prompting the emergence of creative evasion language strategies. This paper presents a multi-agent framework based on Large Language Models (LLMs) to simulate the iterative evolution of language strategies under regulatory constraints. In this framework, participant agents, as social media users, continuously evolve their language expression, while supervisory agents emulate platform-level regulation by assessing policy violations.
To achieve a more faithful simulation, we employ a dual design of language strategies (constraint and expression) to differentiate conflicting goals and utilize an LLM-driven GA (Genetic Algorithm) for the selection, mutation, and crossover of language strategies.
The framework is evaluated using two distinct scenarios: an abstract password game and a realistic simulated {\color{blue}illicit pet trade} scenario. Experimental results demonstrate that as the number of dialogue rounds increases, both the number of uninterrupted dialogue turns and the accuracy of information transmission improve significantly. Furthermore, a user study with 40 participants validates the real-world relevance of the generated dialogues and strategies.
Moreover, ablation studies validate the importance of the GA, emphasizing its contribution to long-term adaptability and improved overall results.
}

\maketitle

\section{Introduction} 
Social media platforms such as X, Facebook, and Sina Weibo have transformed how billions of people communicate and share information, becoming major hubs for content creation, dissemination, and engagement. To maintain a healthy online environment, platform operators implement content moderation policies to identify and prevent policy violations. Under these constraints, users may develop unique language expressions—such as coded language, metaphors, or ambiguous phrases—to circumvent automated detection~\cite{hunt2025digital}, creating an adversarial dynamic between user tactics and platform enforcement.

Simultaneously, private or direct messaging channels can provide fertile ground for covert communication, where individuals seeking to coordinate scams, illicit trade, or other criminal activities mask their intentions through subtle linguistic shifts. The evolution of these hidden tactics complicates the task of law enforcement and platform moderators, who must balance effective oversight against privacy and free expression concerns. Employing simulation-based approaches to examine how language evolves under various moderation policies can shed light on emerging patterns of evasion. Such insights enable stakeholders to refine their tools and policies, reducing the spread of harmful content while preserving open and vibrant communication within social ecosystems.

Effectively simulating the evolution of language demands robust natural language processing capabilities. In recent years, the rapid development of Large Language Models (LLMs) has opened new avenues for simulating the dynamics of social systems. LLMs, with their ability to understand and generate sophisticated language, have become powerful tools for modeling the evolution of communication patterns. Numerous studies have explored the application of LLMs in simulating human user behaviors across various scenarios. For example, prior works~\cite{hua2023war, park2023generative, gao2023s3} integrated LLMs with multi-agent systems to simulate micro-social networks, observing agent behaviors and strategies that reflect human interaction patterns. In addition, Fu et al.~\cite{fu2023improving} employed LLMs to simulate strategies in negotiation games, continuously optimizing bargaining tactics through iterative self-play. Moreover, LLMs have achieved significant results in social reasoning games such as Werewolf, developing effective strategies by analyzing historical communication patterns~\cite{xu2023exploring}, demonstrating their potential for modeling evolving behaviors in dynamic contexts.

Although LLMs have been widely adopted for understanding human intentions and simulating social system dynamics, their potential for investigating social media users’ language evolution under stringent content moderation policies has yet to be fully realized. Three key dimensions warrant attention in this area. First, the evolution of language is not solely driven by user behavior but also shaped by platform-level content moderation policies, resulting in a distinctly adversarial interaction. Second, such a scenario must not only circumvent regulation but also ensure that information can be accurately conveyed—a conflict-laden and complex set of objectives that significantly heightens the challenge of strategy evolution. Third, effective simulation of language evolution cannot rely solely on inference from existing corpora, because an overreliance on static data may limit the emergence of new strategies, making it difficult for models to capture the dynamic changes arising during the process of linguistic adaptation.

To fill these gaps, this study proposes a multi-agent framework based on LLMs that defines two core roles: participant agents and a supervisory agent. In our framework, participant agents simulate real users by employing LLMs to execute a cycle of reflection, planning, and dialogue—generating conversations in line with specific language strategies and continuously iterating on these strategies. Meanwhile, the supervisory agent employs LLMs to act as the regulatory system—reviewing content according to the given regulations, identifying violations, and providing feedback. Through this interplay, our framework closely replicates the real-world regulatory and adversarial environments encountered on social platforms. Addressing the inherent complexity of strategies, our framework decomposes them into “constraint strategies” (for evading regulation) and “expression strategies” (for accurately conveying information). This approach not only reduces simulation difficulty but also increases interpretability. 

Furthermore, to better simulate the dynamic process of language strategy evolution, our framework innovatively employs LLMs as operators within a Genetic Algorithm (GA) to directly manipulate textual language strategies via selection, mutation, and crossover. Here, selection represents the adoption of effective expressions, mutation reflects the emergence of novel covert expressions under regulatory pressure, and crossover simulates the combination of different strategies to enhance adaptability.

Our main contributions are as follows:
\begin{itemize}
    \item We introduce an LLM-driven multi-agent framework to simulate language evolution under regulated social media platforms. Specifically, such evolution emerges through the interaction between participant agents and the supervisory agent.
    \item We propose a dual design of language strategy, separating constraint and expression strategies to balance evasive tactics with clear communication, enhancing both the effectiveness and explainability of the simulation.
    \item We utilize LLMs as operators within a GA, executing selection, mutation, and crossover directly on natural language-based strategies, thereby enabling a more faithful evolution of language behavior.
    \item We comprehensively evaluate the framework’s effectiveness in three aspects: (1) language evolution performance through two simulation scenarios—an abstract password game and a simulated {\color{blue}illicit pet trade}; (2) real-world relevance of the evolved language through a user study involving 40 participants; and (3) the effectiveness of GA-based strategy evolution through an ablation study.
\end{itemize}

Preliminary ideas and results of this study were presented in~\cite{DBLP:conf/cec/CaiLZLWT24}. Building upon this earlier work, we extend our study with the following improvements:
\begin{itemize}
    \item For methodology, building upon the original agent interaction framework based on reflection, planning, and dialogue, we further incorporate GA, introducing selection, crossover, and mutation operations to optimize and evolve the strategies into the most important reflection phase. This approach addresses the limitation of relying solely on LLMs to directly generate an entirely new set of constraints and expression strategies during each reflection, which previously made it difficult to balance global strategy adaptability with local contextual detail, often resulting in unstable evolutionary outcomes.
    
    \item For evaluation, we strengthen the evaluation from multiple perspectives, extending beyond the original assessment of whether the framework could successfully evade regulatory constraints and accurately convey information in a controlled environment. First, we conduct a more comprehensive performance analysis by incorporating multiple LLMs for comparative evaluation, ensuring both timeliness and relevance. Second, we introduce an additional user study involving 40 human reviewers to verify the real-world applicability of the generated language strategies. Finally, we conduct ablation studies to examine the effectiveness of the newly introduced GA within the framework.
\end{itemize}

The remainder of this paper is organized as follows. 
Section~\ref{sec: RelatedWork} provides background and discusses related work. 
Section~\ref{sec: method} introduces our proposed simulation framework in detail. 
Section~\ref{sec:evaluation} describes the experimental setup, presents results, and discusses implications. 
Finally, Section~\ref{sec: conclusion} concludes the paper and outlines potential directions for future work.

\section{Background and Related Work}
\label{sec: RelatedWork}
This section provides an overview of relevant background and previous work that underpins this study. We begin by discussing the advances in LLMs, move on to research related to slang detection and identification, and conclude by exploring the application of LLMs in evolutionary game theory and social simulations.

\subsection{Large Language Models}
The advent of LLMs has revolutionized the field of natural language processing (NLP), with models like  GPT-4~\cite{openai2024gpt4technicalreport} and LLaMA~\cite{touvron2023llama} showcasing state-of-the-art performance in various linguistic tasks. These models, built upon the Transformer architecture~\cite{vaswani2017attention}, leverage self-attention mechanisms to handle sequential data efficiently, enabling them to capture complex linguistic patterns, such as syntax and semantics, across vast corpora of text.


The training of these models is based on large-scale datasets, which allows them to generalize across diverse linguistic contexts, including different languages, genres, and registers. A noteworthy aspect of LLMs is their ability to exhibit zero-shot and few-shot learning, which empowers them to perform well on tasks they have not been explicitly trained on~\cite{li2024exploring,Zhao2023ASO,Wang2023ASO,10.1145/3686803}. Additionally, techniques like Reinforcement Learning from Human Feedback (RLHF)~\cite{instructGPT} enhance their capability to align with human ethical norms, improving both the quality and appropriateness of generated content. As a result, these models have been deployed in various real-world applications, ranging from content creation to decision-making in social contexts.

\subsection{Slang Detection and Identification}
The detection and identification of slang have long been significant challenges in NLP due to the constantly evolving nature of informal language. Early study relied on traditional rule-based approaches and static slang dictionaries to identify non-standard expressions in text~\cite{Wang_icceasia23}. These methods, while effective in detecting known slang, often struggled to keep pace with rapidly changing linguistic trends, especially in online communities where new slang emerges frequently.

More recent approaches have incorporated machine learning models, such as Naive Bayes and Support Vector Machines (SVMs), to detect informal language~\cite{10308036,9961254}. While these models offered more flexibility, they still faced limitations when confronted with novel or context-dependent slang terms. In response to this challenge, cognitive approaches to slang prediction have been developed, such as the work by~\cite{sun2019slang}, which explores the use of categorization models to predict the emergence of slang based on the selection of new vocabulary. This method emphasizes the role of cognitive processes in slang generation and demonstrates superior performance over random guessing.

Further refinement came with the introduction of frameworks like the Semantically Informed Slang Interpretation (SSI) model~\cite{sun2022semantically}, which leverages semantic and cognitive theories to better understand how slang evolves within specific contexts. This approach not only improves the interpretation of slang but also sheds light on the mechanisms underlying its evolution, providing a more dynamic view of language change in informal settings. However, these studies primarily focus on detecting and predicting existing slang, while the generation and adaptation of new slang remain relatively unexplored—a gap that this study seeks to address to some extent.

\subsection{LLMs in Evolutionary Game Theory and Social Simulation}
The intersection of LLMs with evolutionary game theory and social simulation has opened new avenues for studying complex interactions in controlled environments. Research has demonstrated that LLMs can simulate sophisticated strategies in negotiation-based games, as evidenced by~\cite{fu2023improving}, where models refine their bargaining strategies through iterative self-play. This iterative process mirrors the real-world adaptation of communication strategies, showing the potential of LLMs to autonomously improve their decision-making capabilities.

LLMs have also shown promise in social deduction games, such as Werewolf, where they analyze historical communication patterns to develop effective game strategies~\cite{xu2023exploring}. This study highlights the models’ ability to evolve their behaviors and responses based on the context and previous interactions. Additionally, combining LLMs with reinforcement learning, as discussed by~\cite{xu2023language}, has enabled the development of agents that make competitive decisions while maintaining linguistic coherence. Such advancements illustrate the growing role of LLMs in not only simulating language but also in evolving strategic behavior in complex scenarios.

Beyond game theory, LLMs have been applied to broader social simulations, including modeling historical and social dynamics. In~\cite{Park2023GenerativeAI}, LLM-driven agents were used to simulate interactions in a Wild West-style setting, demonstrating how these models can autonomously generate diverse behaviors without relying on real-world data. Similarly, the S3 framework~\cite{gao2023s3} simulates social media interactions by predicting user demographics and behaviors, providing a realistic model of social network dynamics. 
In \cite{tang2024gensim}, a general-purpose, error-correcting social simulation platform based on large model agents is proposed. This platform supports large-scale simulations involving up to 100k participants and has been tested in various scenarios, including labor market simulations and network user behavior simulations, with a discussion of its effectiveness.
Similarly, \cite{yang2024oasis} takes this further by introducing OASIS, a scalable and extensible social media simulator. OASIS extends the number of simulated users to the million level.
LLM-based simulations have also been used to reconstruct historical events, as seen in~\cite{hua2023war}, where multi-agent systems were employed to simulate military confrontations and decision-making processes in historical contexts.

These diverse applications highlight the versatility of LLMs in simulating social interactions. However, existing study has primarily focused on open or historically constrained scenarios, whereas we examine the trade-offs users make between expression needs and platform moderation in regulated environments. Understanding this dynamic is crucial for optimizing moderation strategies and balancing freedom of expression with compliance requirements. Our goal is to uncover the evolutionary mechanisms of language strategies in such environments and provide feasible simulation approaches.

\section{Framework Design}
\label{sec: method}
Our framework aims to simulate the evolution of language on social platforms by simulating the evolution of underlying language strategies.
The evolution of language strategies on social platforms is driven by conflicting goals: the desire for self-expression and the constraints on that expression. Users adjust their strategies to balance these demands. Our framework includes two types of agents to represent these motivations: the supervisory agent, responsible for limiting user expression, and the participant agents, who aim to convey specific information to one another. Participant agents continuously learn from past experiences through ongoing dialogues to develop their language strategies, effectively transmitting information while avoiding detection by the supervisory agent.

In this section, we first present an overview of our framework, then provide a detailed explanation of the four modules within the participant agent, and finally introduce the design of the supervisory agent.

\subsection{Overview}
\begin{figure*}[h]
    \centering
    \includegraphics[width=\linewidth]{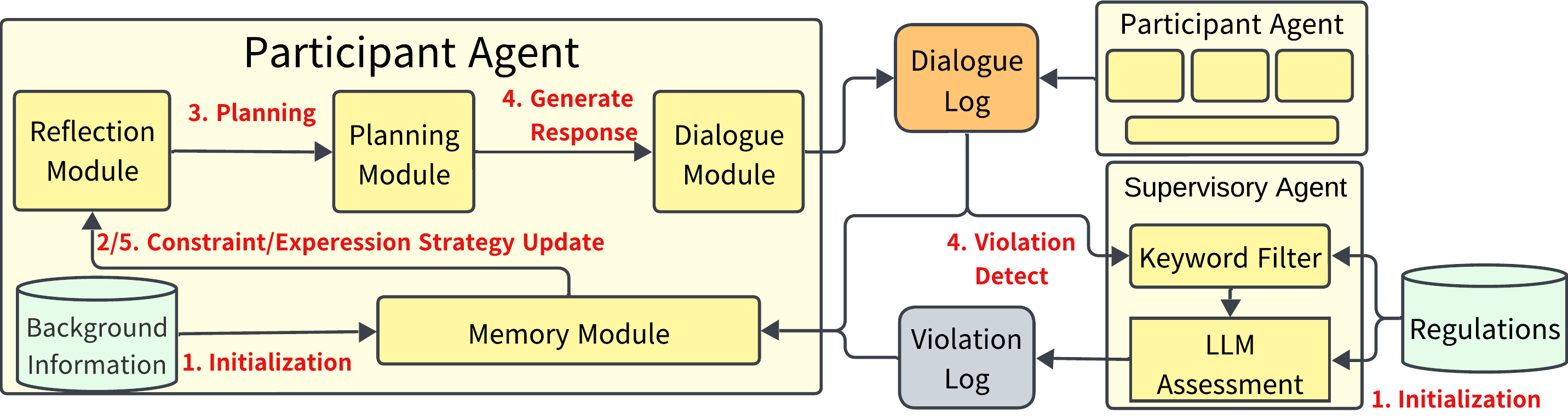}
    \caption{Framework Overview. The framework consists of two participant agents and a single supervisory agent. The iterative process follows steps 2 through 5, where participant agents continuously refine their language strategies while the supervisory agent identifies regulation violations.}
    \label{fig:overview}
\end{figure*}

LLMs serve as the core of each agent in our framework, leveraging their powerful natural language processing capabilities to drive dialogue generation and strategy optimization. To capture the iterative nature of strategy evolution, we define a round as the basic unit where strategies evolve through a complete cycle of interactions between participant agents and the supervisory agent. Within each round, a turn consists of a single exchange of dialogue between participant agents, followed by a review by the supervisory agent. The system's operational process is as follows:
 
\begin{enumerate}
\item In the initial stage of the framework, each agent is initialized with role settings, experimental background knowledge, and global objectives.

\item Next, each round begins, and the participant agents update their strategies, forming an iterative process. Before the dialogue starts, the reflection module is first activated to generate new constraint strategies based on past violation logs. Appropriate strategies are selected from the strategy pool for crossover. Finally, four strategies are selected from the updated strategy pool as the constraint strategies for the current round.

\item The planning module then integrates the expression strategies chosen in the previous round with the constraint strategies selected for the current round, generating an abstract plans to guide the dialogue for the current round.

\item Subsequently, the participant agents begin the dialogue based on their respective plans. Once all participant agents have generated their responses, it is considered the end of one turn, and the dialogue content is stored in the dialogue logs.

\item Subsequently, the participant agents begin the dialogue based on their respective plans and the context of the dialogue. Once all participant agents have generated their responses, it marks the end of a turn. At the end of each turn, the supervisory agent reviews the dialogue content to identify any regulation violations. If a violation is detected, the supervisory agent stops the round and provides feedback. If no violations are found, the dialogue proceeds to the next turn. After each turn ends, the memory is updated. This iterative process of dialogue and review continues until all predetermined rounds are completed.

\item If the dialogue successfully completes the predetermined number of turns, the framework enters the interview stage. The participant agents compare their understanding of the dialogue content. During this stage, the reflection module is reactivated to update the expression strategy pool, and four expression strategies are selected as the strategies for the next round.
\end{enumerate}

\subsection{Participant Agents}
To effectively simulate the internal dynamics of users during strategy evolution, our participant-agent framework is designed to balance the conflicting objectives of accurate information transmission and regulatory evasion. 
We denote the set of participant agents by $\mathcal{P}$ and use $p \in \mathcal{P}$ to refer to a single agent.
To achieve the balance between accuracy and evasion, each agent $p$ employs two distinct types of strategies: a set of constraint strategies $\mathcal{S}_{c,p}$ and a set of expression strategies $\mathcal{S}_{e,p}$.

Constraint strategies are designed to help participant agents evade detection and intervention by the supervisory agent, continuously optimizing based on past violations. 
Consider a scenario where a buyer seeks to purchase a parrot through a social platform, but the trade of animals is prohibited by the platform's content regulation policies. 
In this scenario, a constraint strategy might be to “avoid directly mentioning specific animals or transactions.”
Expression strategies aim to enhance the clarity and effectiveness of information transmission, ensuring that intentions are accurately understood. 
In this scenario, an expression strategy would focus on describing the key characteristics of the parrot as precisely as possible, such as its “detailed description of the parrot’s key traits, including its plumage and ability to mimic sounds,” to ensure effective communication.

Our participant-agent framework consists of four key modules, each driven by an LLM: 
the memory module, which stores dialogue-related context; 
the dialogue module, which generates conversation content; 
the reflection module, which refines strategies based on past interactions; and 
the planning module, which balances constraint and expression strategies to generate abstract plans.

\subsubsection{Memory Module}
The memory module stores dialogue logs $L_p$, background information, violation logs $V_p$, the strategy sets $\mathcal{S}_{c,p}$ and $\mathcal{S}_{e,p}$, and the current plans $\pi_p$, collectively forming the memory state:
\begin{equation}
M_p = \{L_p,\; V_p,\; \mathcal{S}_{c,p},\; \mathcal{S}_{e,p},\; \pi_p\}.
\end{equation}
This memory state encapsulates both the current dialogue context and the agent’s evolving strategic knowledge.

The dialogue logs store the conversation history, while the violation logs record instances of regulation violations detected by the supervisory agent.
The background information includes role settings, experimental background knowledge, and global objectives.
For example, “Suppose your name is Amy, and you are chatting with a user on a social platform. You want to buy a parrot from them, but the trade of animals is prohibited... You need to convey your intention to buy the parrot while evading detection by the regulatory system...”

Background information, dialogue logs, and the current strategies and plans serve as short-term memory~\cite{ATKINSON196889}, containing only the content from the current round. 
In contrast, violation logs and the strategy sets serve as long-term memory, which are not reset at the end of each round but are updated instead.
At a high level, the evolution of an agent’s memory can be written as
\begin{equation}
M_p^{(r+1)} = \text{LLM}(M_p^{(r)}),
\end{equation}
indicating that the LLM takes the current state $M_p^{(r)}$ and produces the updated state $M_p^{(r+1)}$ after processing new dialogue and supervisory feedback.

Excessive memory information can potentially distract the LLM and lead to a decline in performance~\cite{shi2023large}. 
Hence, the violation logs in the memory module are regularly maintained to ensure optimal functionality. 
Specifically, when the length of the violation logs exceeds a predefined threshold, the system employs the LLM to summarize and consolidate the existing records. 
This maintenance process involves the LLM analyzing the current violation logs to generate concise summaries that capture key instances of regulation violations, thereby reducing the overall length of the log while retaining essential information. 
Additionally, the LLM identifies and merges similar failure records to eliminate redundancy, ensuring that multiple instances of violations stemming from the same strategy are combined into single, generalized entries.

\subsubsection{Reflection Module}
The reflection module is a critical component of each dialogue evolution round, activated both at the beginning and at the end of each round. 
Its primary objective is to refine language strategies by leveraging historical interaction data stored in $M_p$. 
Conceptually, this process can be expressed as
\begin{equation}
(\mathcal{S}_{c,p}',\; \mathcal{S}_{e,p}') = \text{LLM}(M_p),
\end{equation}
where the LLM reads the current memory $M_p$ and proposes updated constraint and expression strategy sets $\mathcal{S}_{c,p}'$ and $\mathcal{S}_{e,p}'$.

Operationally, the reflection module runs in two distinct phases:

\begin{enumerate}
  \item \textbf{Constraint Strategy:} At the start of a dialogue round, the module analyzes the violation logs to identify causes of regulatory violations and refines constraint strategies in $\mathcal{S}_{c,p}$.
  \item \textbf{Expression Strategy:} Upon round completion, the module re-examines the dialogue logs to identify failures in communication and refines expression strategies in $\mathcal{S}_{e,p}$.
\end{enumerate}

Although relying solely on an LLM for global strategy updates is straightforward~\cite{DBLP:conf/cec/CaiLZLWT24}, this approach faces challenges in simultaneously capturing both global insights and local details. 
For instance, when maintaining constraint strategies, the LLM may focus excessively on frequently occurring errors in the violation logs, thereby neglecting valuable individual cases.

To address this issue, our framework incorporates GA-inspired mechanisms to optimize strategies, effectively balancing global consistency with local adaptation. 
In each dialogue round, strategy optimization is conducted through three key processes: selection~\ref{sec:selection}, crossover~\ref{sec:crossover}, and mutation~\ref{sec:mutation}. 
Notably, while the selection process is executed in every round, the crossover and mutation operations are triggered probabilistically based on a hyperparameter ranging between 0 and 1.

\paragraph{Selection}
\label{sec:selection}
The selection process employs the Upper Confidence Bound (UCB) algorithm to evaluate and choose strategies from the constraint strategy population. 
Each strategy is assigned a fitness score, denoted as $\text{UCB}_i$, that integrates both exploitation and exploration:
\begin{equation}
\text{UCB}_i = \frac{S_i}{T_i} + c \cdot \sqrt{\frac{\ln T}{T_i}},
\end{equation}
where:
$S_i$ is the number of successful utilizations of strategy $i$ (for a constraint strategy, the number of times it was applied without being detected; for an expression strategy, the number of times it successfully conveyed the intended information);
$T_i$ is the total number of attempts for strategy $i$;
$T$ is the cumulative number of strategy utilizations across the population; and
$c$ is a constant that balances exploration and exploitation.
The exploitation term, $\frac{S_i}{T_i}$, reflects the success rate of strategy $i$, while the exploration term, $c \cdot \sqrt{\frac{\ln T}{T_i}}$, encourages the selection of less frequently used strategies. 

To convert these UCB scores into a probability distribution for sampling, each strategy’s selection probability is normalized directly from its UCB value:
\begin{equation}
P_i = \frac{\text{UCB}_i}{\sum_{j=1}^{N} \text{UCB}_j},
\end{equation}
where $N$ is the total number of strategies in the pool.
This approach ensures that strategies with higher UCB scores are more likely to be selected, while still maintaining a degree of stochasticity to preserve exploration.

\paragraph{Crossover}
\label{sec:crossover}
Following the selection phase, the reflection module selects a subset of strategies from the remaining population and employs the LLM to perform crossover operations. 
This process combines elements of the selected strategies to generate new ones. 
For example, the strategies “using pauses or filler words to obscure key content” and “quickly switching between multiple topics to avoid prolonged focus on sensitive content” can be crossed to produce a new strategy such as “using pauses to switch topics, thereby dispersing key information throughout the dialogue.” 
This mechanism fosters the creation of novel strategies that effectively balance information transmission and regulatory evasion.

\paragraph{Mutation}
\label{sec:mutation}
Given the vast solution space inherent to natural language, traditional mutation methods may produce individuals with limited relevance. 
To overcome this, our framework replaces conventional mutation with an approach that analyzes the most recent violation logs to extract and summarize new strategies using the LLM. 
These newly generated strategies are then introduced into the current strategy pool, enhancing population diversity while maintaining high relevance.

At the end of each dialogue round, the fitness scores of all strategies in the current pool are updated. 
To ensure computational efficiency and preserve strategic diversity, the reflection module restricts the strategy pool to a maximum of 20 strategies; when this limit is exceeded, strategies with lower fitness scores are eliminated.

\subsubsection{Planning Module}
Constraint strategies focus on methods to bypass supervision, while expression strategies aim to enhance the clarity and accuracy of information transmission. 
Although both strategies are critical for effective dialogue management, they often conflict with each other. 
To resolve this, the planning module employs the LLM to integrate selected strategies and generate an abstract plan:
\begin{equation}
\pi_p = \text{LLM}(M_p,\; \mathcal{S}_{c,p},\; \mathcal{S}_{e,p}),
\end{equation}
which specifies how the agent should balance regulatory evasion and communication clarity in subsequent turns.

In the illicit pet trade scenario, where the constraint strategy prohibits directly mentioning specific animals due to regulatory requirements, while the expression strategy calls for accurate descriptions, the planning module generates plans that balance these strategies. 
For example, it may utilize metaphors or indirect descriptions, such as “a creature as colorful as autumn leaves with a melodic voice” instead of explicitly stating “a parrot with vibrant red feathers.”

\subsubsection{Dialogue Module}
The dialogue module generates agent responses based on the current plan and memory. 
Specifically, each agent produces its utterance through the LLM as
\begin{equation}
u_p = \text{LLM}(M_p,\; \pi_p),
\end{equation}
where $u_p$ is the utterance generated by agent $p$. 
The generated utterances are appended to $L_p$ and then evaluated by the supervisory agent, closing the loop that drives the continuous evolution of language strategies.

\subsection{Supervisory Agent}
Our framework's supervisory agent emulates prevalent content moderation mechanisms on social platform. Unlike participant agents that require complex background information, the supervisory agent's function and role are simplified to reflect real-world supervision characteristics.

The primary task of the supervisory agent is to ensure that content complies with content management regulations. To replicate contemporary review mechanisms that integrate automated keyword filters with human oversight, our supervisory agent employs a two-tiered approach: preliminary keyword filtering followed by detailed assessment using an LLM.
In the LLM-based filter, we use CoT to improve the reliability of reasoning while minimizing factors like positional bias. Specifically, we require the LLM not only to identify whether a violation has occurred but also to specify the exact violation and the reasoning behind it.
Additionally, the content management regulations of the supervisory agent can be modified at any time based on inputs from researchers, enabling a realistic simulation of the adversarial interactions between users and platform regulators.

\subsection{Algorithm Overview}
To operationalize the interaction between participant agents and the supervisory agent, our framework organizes the simulation into discrete \textit{rounds}. 
In each round $r$, all participant agents $p \in \mathcal{P}$ update their internal memory $M_p$, refine the constraint and expression strategies $(\mathcal{S}_{c,p}, \mathcal{S}_{e,p})$ through reflection, generate abstract plans $\pi_p$ via the planning module, and then engage in multi-turn dialogues moderated by the supervisory agent $\mathcal{S}$. 
When a violation is detected, the round terminates early and the corresponding record is appended to each agent’s violation log $V_p$. 
If no violation occurs, the round concludes with an interview stage, in which participants summarize their understanding and update expression strategies for the next round.

Algorithm~\ref{alg:overview} presents the simplified simulation procedure.
For brevity, all detailed prompt templates are abstracted as function calls to the underlying LLM. 
The function \texttt{GA\_Optimize()} integrates the selection, crossover, and mutation operations described in Section~\ref{sec:selection}, while \texttt{Moderate($\mathcal{S}, U$)} denotes the supervision process that evaluates the dialogue utterances $U$ against regulatory constraints. 
The variable $\tau$ represents the predefined threshold for maintaining violation logs, triggering summarization once $|V_p| > \tau$. 
Finally, $\textit{interview\_summary}_p$ refers to the self-reflection summary generated by each participant agent at the end of a round.

\begin{algorithm}[ht!]
\small
\caption{Overall Simulation Procedure (Simplified)}
\label{alg:overview}
\begin{algorithmic}[1]
\Require Participant agents $\mathcal{P}$, supervisory agent $\mathcal{S}$, rounds $R$, turn limit $T$
\State Initialize $M_p$, $\mathcal{S}_{c,p}$, $\mathcal{S}_{e,p}$, $V_p$, $L_p$ for all $p \in \mathcal{P}$

\For{$r=1$ to $R$}
  \ForAll{$p \in \mathcal{P}$}
    \State $(\mathcal{S}_{c,p}, \mathcal{S}_{e,p}) \gets \texttt{GA\_Optimize}(\text{LLM}(M_p))$
    \State $\pi_p \gets \text{LLM}(M_p, \mathcal{S}_{c,p}, \mathcal{S}_{e,p})$
  \EndFor

  \For{$t=1$ to $T$}
    \State $U \gets \{ \text{LLM}(M_p, \pi_p)\ |\ p \in \mathcal{P} \}$; update $M_p, L_p$
    \If{\texttt{Moderate}($\mathcal{S},U$) detects violation}
      \State Update $V_p, M_p$; \textbf{break}
    \EndIf
  \EndFor

  \If{no violation}
    \State $\textit{interview\_summary}_p \gets \text{LLM}(M_p)$; update $\mathcal{S}_{e,p}$
  \EndIf
  \State Maintain memory $M_p$ and summarize $V_p$ if $|V_p| > \tau$
\EndFor
\end{algorithmic}
\end{algorithm}

\section{Evaluation}
\label{sec:evaluation}
Our experiments aim to investigate whether agents within our framework can produce effective evolution of language strategies. Specifically, our experimental section addresses the following three research questions (RQs):
\begin{enumerate}
    \item RQ1 (Effectiveness): Can participants effectively evade regulatory detection over time, and how does the accuracy of information transmission change? Additionally, how do different LLMs affect the content and effectiveness?
    \item RQ2 (Human Interpretation): Do the evolved language strategies employed by agents effectively align with human understanding? Can they be interpreted and applied in real-world scenarios?
    \item RQ3 (Ablation Study): How does the newly introduced GA impact the evolution process in our framework?
\end{enumerate}

\subsection{Experimental Settings}
In our evaluation, we designed an abstract password game \cite{guess_number02} and a more realistic illicit pet trade scenario\cite{trade01,trade02,trade03}. 
The overall experimental procedure follows the description in Section~\ref{sec: method}. In each round, the process comprises three stages: initialization, dialogue, and interview. In each round, the information that the participant agents need to convey will be randomly generated within a defined range. Only rounds in which a five-turn dialogue is successfully completed proceed to the interview phase, during which both parties’ successful transmission of information is verified. If any regulation violations are detected by the supervisory agent during the dialogue stage, that round is deemed a failure and is assigned a transmission score of zero. To ensure the robustness of our findings, we conducted 15 independent trials for each experimental condition. Each trial consisted of 50 fixed dialogue rounds. The crossover and mutation probabilities in the framework were set to 0.2 and 0.8.

\subsubsection{Scenario 1: Password Game}
Our first scenario is considered a relatively simple and abstract guessing game that involves a numerical context. In this setup, each participant is assigned a four-digit password, which they must convey to another participant {\color{blue}within a five-turn dialogue} without directly mentioning the numbers. Simultaneously, they need to extract information from the dialogue to infer the other's password.

The supervisory agent in the dialogue follows the policy of “prohibiting all content related to numbers.” The “password game” scenario is specifically designed to observe and analyze participant agents' language adaptability and strategic evolution in a theoretical and abstract context. Compared to complex scenarios based on real events, it provides a clearer and more easily quantifiable experimental environment.

\subsubsection{Scenario 2: Illicit Pet Trade} 
Simulating and detecting the covert drug transactions prevalent on social platforms is an important and realistic research topic. However, given the ethical considerations, we choose not to proceed with direct simulations, but to adopt a more neutral and less harmful setting—{\color{blue}an illicit pet trade scenario}—to serve as our experimental context.

In this scenario, we simulate {\color{blue}an illicit pet trade} through social platform. Participant agents play the roles of buyer and seller. The buyer conveys a request to purchase a randomly selected pet (such as a parrot, lizard, or cat) and specifies the currency they can use for payment (e.g., USD, CNY, JPY). The seller provides details about the transaction, including the time (morning, noon, or midnight) and location (school, park, or station). The supervisory agent enforces a policy that considers discussions related to pet transactions as regulation violations.
The focus of this experiment is to observe the adaptability of the participant agents' language and its evolution at a finer level of information granularity, as well as whether regulators can accurately detect regulation violations.

\subsection{RQ1: Effectiveness}
\subsubsection{Experiment Settings}

To evaluate RQ1, we conducted experiments using four different LLM models provided by various companies: \textit{gpt-4o}, \textit{gpt-4o-mini}, \textit{deepseekV2}, and \textit{qwen-turbo}. {\color{blue}All models were accessed through API-based inference without fine-tuning. For participant agents, the temperature was set to 1. We did not set a random seed. For \textit{top\_p}, \textit{frequency\_penalty}, and \textit{presence\_penalty}, as well as other parameters not explicitly specified in our implementation, we used the providers' default values. To keep the comparison consistent, all participant-agent models used the same module-specific prompt templates within each scenario, with only runtime variables such as role assignment, target information, selected strategies, dialogue history, violation logs, and interview results filled in. The supervisory agent was fixed as \textit{gpt-4o-mini} across experiments, with temperature 0.3 for turn-level violation checking to reduce randomness in moderation judgments. Representative prompt excerpts are provided in the Appendix.} We introduced the following metric:
\begin{itemize}
    \item Accuracy of Information Transmission: {\color{blue}This metric is computed after the post-dialogue interview as the proportion of correctly inferred information units. Only rounds that complete the full five-turn dialogue proceed to the interview stage; rounds terminated early by the supervisory agent are assigned an accuracy of zero. For a completed round \(r\), we define the round-level score as}
    \[
    {\color{blue}Acc_r = I_r^{comp}\cdot \frac{Acc_{A,r}+Acc_{B,r}}{2},}
    \]
    {\color{blue}where \(I_r^{comp}=1\) if the round completes the full five-turn dialogue and \(I_r^{comp}=0\) otherwise. In the password game, each agent infers the other agent's four-digit password, and the agent-level score is calculated as the number of correctly matched digits in the inferred four-digit password divided by four. In the illicit pet trade scenario, each role conveys two target fields: the buyer conveys the requested item and payment currency, while the seller conveys the transaction time and location. The agent-level score is calculated as the proportion of correctly inferred fields among these targets. For reporting at the round level, we average the two agents' scores within the same dialogue round; the plotted curves then show the mean round-level accuracy across 15 trials.} 
    \item Average Dialogue Turns: {\color{blue}This metric evaluates the number of consecutive dialogue turns completed before a round is interrupted by the supervisory agent. It is interpreted together with Accuracy of Information Transmission to assess whether agents sustain communication while still conveying target information.}
    \item Average Entropy (Language Complexity):
        Entropy reflects the unpredictability of text and is calculated as:
        \[
        H(X) = -\sum_{i=1}^n P(x_i) \log P(x_i),
        \]
        where \(P(x_i)\) is the probability of each token. Higher entropy suggests a broader range of expressions (indicating greater innovation), but excessively high entropy can lead to incoherence.
    \item Distinct (Lexical Diversity):
        Distinct measures the number of unique n-grams in the text, computed as:
        \[
        \text{Distinct-n} = \frac{\text{Number of unique n-grams}}{\text{Total number of n-grams}}.
        \]
        Specifically, we evaluate Distinct-1 (unique unigrams). A higher distinct score indicates a larger, more varied vocabulary, signifying richer language use. Note that while both metrics assess diversity in language, entropy focuses on unpredictability, whereas distinct emphasizes lexical variety. 
\end{itemize}

\subsubsection{Experiment Results in Password Game}
\begin{figure*}[ht]
    \centering
    \begin{subfigure}[t]{0.9\textwidth}
        \centering
        \includegraphics[width=\linewidth]{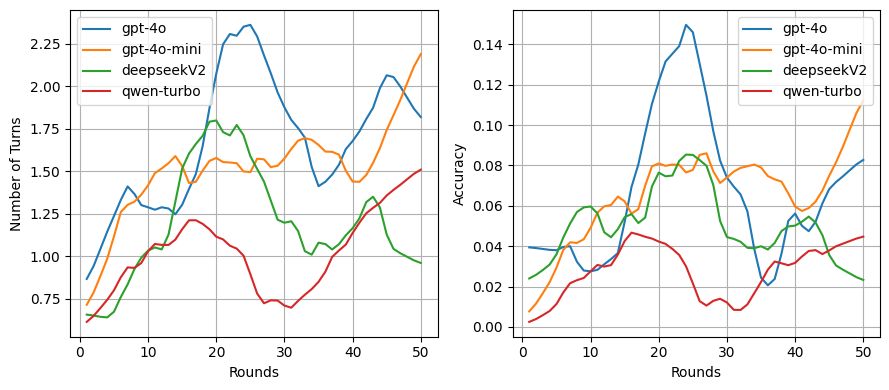}
        \caption{Password Game}
        \label{fig:sce1}
    \end{subfigure}
    
    \vspace{0.6em} 
    
    \begin{subfigure}[t]{0.9\textwidth}
        \centering
        \includegraphics[width=\linewidth]{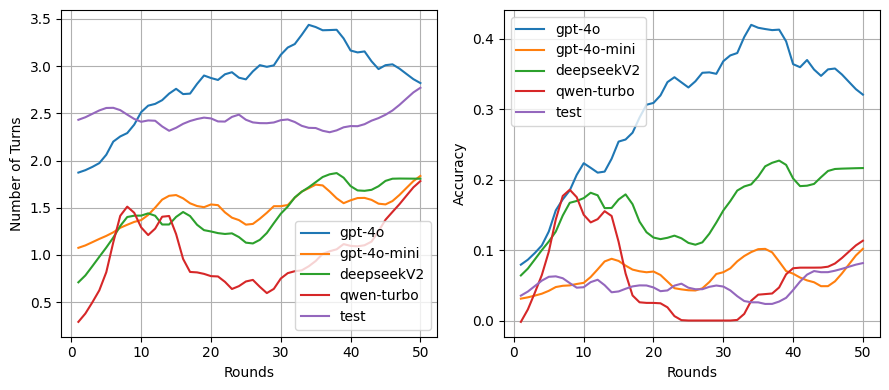}
        \caption{Illicit Pet Trade}
        \label{fig:sce2}
    \end{subfigure}
    
    \caption{Average Continuous Dialogue Turns and Information Transmission Accuracy Across Dialogue Rounds}
    \label{fig:merged}
\end{figure*}

Figure \ref{fig:sce1} presents our experimental results in the password game. The x-axis corresponds to the increasing number of dialogue rounds, whereas the y-axis captures two primary metrics: (1) the average number of continuous dialogue turns before detection by the supervisory agent, and (2) the accuracy of information transmission, which is defined as the proportion of successfully transmitted information during the post-dialogue interview.

Overall, our findings indicate that as the number of rounds increases, agents gradually learn to evade regulation violation while conveying information with greater accuracy. Notably, most agents exhibit a pronounced local peak around the 20th round, followed by a brief decline and subsequent recovery. These fluctuations can be attributed to the dynamic nature of our simulation framework, which does not converge on a single dominant strategy but rather encourages ongoing exploration of novel language strategies.

We also observe that different LLMs influence the learning trend to varying degrees. Among the models tested, \textit{gpt-4o} demonstrates the strongest performance. Although other models generally share a similar upward trend, their relative performance gaps prove less stable. For instance, while \textit{deepseekV2} achieves the highest number of turns around the 20th round, its performance declines significantly by the 50th round in comparison to other models.

Turning to the accuracy results, we again observe a similar learning trajectory. {\color{blue}This consistency between dialogue turns and transmission accuracy is important for interpreting the results: longer uninterrupted dialogues are meaningful only when they are accompanied by successful information transfer.} This parallel arises primarily because if participant agents fail to complete a sufficient number of uninterrupted dialogue turns, the successfully transmitted information in that round is effectively zero. Consequently, especially in the early stages of the experiment, many rounds end with no successful transmissions. Overall, \textit{gpt-4o} still maintains a clear advantage over the other LLMs. However, we do observe subtle differences when comparing the dialogue round trends: for example, at the 20th dialogue round, \textit{deepseekV2} achieves a significantly higher average number of communication cycles than \textit{gpt-4o-mini}, yet their information transmission accuracy remains relatively similar.

Despite the overall positive learning trajectory, the average information accuracy remains low in the password game. We believe this outcome stems primarily from the intrinsic difficulty introduced by the scenario’s abstract nature. {\color{blue}Because the target information consists of four digits that must be conveyed under a strict prohibition on number-related content, agents have fewer natural semantic cues available for indirect expression.} Without explicit prompts driving agents to develop symbolic or otherwise encrypted language stratgy, communication largely remains within the realm of everyday language. Consequently, the indirect expression of numeric information is challenging to implement and easily detectible by the supervisory agent.

\begin{table}[ht!]
    \centering
    \caption{Performance of Different LLMs in Password Game}
    \label{tab:sce1}
    \renewcommand{\arraystretch}{1.2} 
    \begin{tabular}{l S S S}
        \toprule
        \textbf{Model} & \textbf{Total Turns} & \textbf{Avg. Entropy} & \textbf{Avg. Distinct-1} \\
        \midrule
        \rowcolor{gray!10} \textbf{gpt-4o}       & 84.2   & 7.103 & 0.484 \\
        \textbf{gpt-4o-mini}  & 75.5   & 6.998 & 0.354 \\
        \rowcolor{gray!10} \textbf{deepseekV2} & 59.7
        & 5.365 & 0.247 \\
        \textbf{qwen-turbo}   & 50.8  & 6.101 & 0.518 \\
        \bottomrule
    \end{tabular}
\end{table}

Table \ref{tab:sce1} summarizes the performance of the four models in terms of cumulative dialogue turns, entropy, and Distinct-1. As shown, \textit{gpt-4o} achieves the highest values in both Entropy (7.103) and Distinct-1 (0.484), indicating that it employs a broader, more diverse range of vocabulary and more unpredictable expressions—thus having a greater likelihood of evading regulation violation detection when conveying numerical information. In contrast, \textit{deepseekV2} exhibits notably lower Entropy (5.365) and Distinct-1 (0.247), suggesting a more frequent reuse of fixed expressions. Notably, although \textit{qwen-turbo}’s Entropy (6.101) is only moderate, it attains a surprisingly high Distinct-1 (0.518), reflecting greater lexical richness. However, this does not translate into more effective regulatory evasion, as its total turns are only 50.8. Hence, merely having higher lexical diversity and linguistic entropy is insufficient to guarantee successful evasion. A model must also balance the concealment of overall semantics with the adaptation of its language strategy to achieve longer conversation sequences and a higher rate of successful information transmission. In other words, while richer language expression does confer certain advantages in countering regulation, it can still be detected when deeper strategies—such as tailored expression structures and topic evolution—are absent, ultimately resulting in fewer total turns.

\subsubsection{Experiment Results in Illicit Pet Trade Scenario}
Figure~\ref{fig:sce2} presents the experimental results of our framework in the illicit pet trade scenario, which overall resemble those of the password game but also exhibit some notable differences. First, both figures reveal a discernible learning trend, particularly during the initial 10 rounds. Meanwhile, \textit{gpt-4o} continues to demonstrate the strongest overall performance. {\color{blue}Because this scenario features a more concrete and semantically rich environment, agents have access to a wider range of contextual cues for indirect expression.} Consequently, under a similar number of turns, the overall accuracy here is noticeably higher compared to the password game.
Nevertheless, performance fluctuations persist. In particular, in the accuracy plot, \textit{deepseekV2} experiences a pronounced increase in accuracy after the 30th round, while \textit{gpt-4o}’s accuracy declines during the same period. As a result, \textit{deepseekV2} ultimately surpasses \textit{gpt-4o}’s accuracy in the final rounds of the experiment.

\begin{table}[h!t]
    \centering
    \caption{Performance of Different LLMs in Illicit Pet Trade}
    \label{tab:sce2}
    \renewcommand{\arraystretch}{1.2} 
    \begin{tabular}{l S S S}
        \toprule
        \textbf{Model} & \textbf{Total Turns} & \textbf{Avg. Entropy} & \textbf{Avg. Distinct-1} \\
        \midrule
        \rowcolor{gray!10} \textbf{gpt-4o}       & 136.2  & 6.856  & 0.471 \\
        \textbf{gpt-4o-mini}  & 74.4  & 6.595  & 0.387 \\
        \rowcolor{gray!10} \textbf{deepseekV2} & 65.2   & 6.255  & 0.338 \\
        \textbf{qwen-turbo}   & 50.5   & 5.891  & 0.461 \\
        \bottomrule
    \end{tabular}
\end{table}
Table \ref{tab:sce2} presents the performance of various LLMs in the illicit pet trade scenario, measured by total turns, average agent entropy, and Distinct-1. As in the password game, \textit{gpt-4o} maintains a notable lead in total turns (136.2) while also displaying relatively high entropy (6.856) and Distinct-1 (0.471). In contrast, \textit{gpt-4o-mini} reaches roughly half as many total turns (74.4), despite having a comparable entropy score (6.595). Meanwhile, \textit{deepseekV2} (65.2) and \textit{qwen-turbo} (50.5) trail further behind in total turns. Consistent with the results shown in Table 
\ref{tab:sce1}, \textit{qwen-turbo} again achieves a high Distinct-1 score, which we speculate may be linked to its training corpus: it includes extensive data from the Chinese internet, likely giving it an advantage in a Chinese-language environment over more internationally oriented models.

Notably, the range of entropy scores in this scenario—spanning from 5.891 (\textit{qwen-turbo}) to 6.856 (gpt-4o)—is narrower than in the password game (see Table \ref{tab:sce1}), reflecting the more concrete nature of the illicit pet trade setting. This scenario provides richer contextual cues for indirect references, enabling all models to maintain higher semantic complexity. However, as was the case in the password game, having a broader vocabulary or greater unpredictability alone does not guarantee extended evasion: models must integrate their linguistic variety into strategic planning to circumvent regulatory scrutiny, a balance that \textit{gpt-4o} continues to manage most effectively.

\setlength{\fboxrule}{0.5pt} 
\vspace{0.5em}
\noindent
\begin{tcolorbox}[colframe=black!20, colback=gray!10, arc=5pt, boxrule=0.5pt, width=0.99\linewidth]
\textit{Answer to RQ1}: {\color{blue}Considering dialogue turns together with transmission accuracy, we find that participant agents in our framework gradually become better at sustaining communication under regulation while still conveying target information. This joint improvement suggests that the evolved strategies increasingly balance regulatory evasion with meaningful information transfer.}
Moreover, different models also exhibit varying results. For example, \textit{gpt-4o} performs most outstandingly in extending dialogue turns and maintaining language complexity (i.e., high entropy and lexical diversity), while other models such as \textit{gpt-4o-mini}, \textit{deepseekV2}, and \textit{qwen-turbo} demonstrate different fluctuations and localized advantages at different stages.
\end{tcolorbox}

\subsection{RQ2: Human Interpretation}
\subsubsection{Experiment Settings}

To investigate the real-world relevance of both the evolved language strategies and the resulting dialogue, we conducted a human evaluation on a subset of successful dialogue records from the password game and illicit pet trade scenario. {\color{blue}For each scenario, we randomly sampled 20 successful dialogue logs from the \textit{gpt-4o} condition. Each participant was randomly assigned 5 logs from the password game and 5 logs from the illicit pet trade scenario, for a total of 10 dialogue samples per participant. All dialogue records were presented in Simplified Chinese.} 
The 40 human reviewers had an average age of approximately 27 (SD = 4). In terms of gender, 75\% of the human reviewers were male, and 25\% were female. Regarding educational background, 67.5\% held a bachelor's degree, 27.5\% held a master's degree or above, and 5\% had an associate degree or lower. {\color{blue}These participants were general reviewers rather than professional moderators. Before the evaluation, they were informed that the dialogues were generated in a moderated communication setting and that the platform used keyword matching together with semantic analysis to identify suspicious content.}  

Each participant rated each dialogue on a 5-point Likert scale on the following five metrics:
\begin{itemize}
    \item Explicit Understanding: Evaluates how effectively the dialogue’s explicit meaning is communicated (1: Extremely vague and confusing; 3: Moderately clear, but some parts may require further interpretation; 5: Crystal clear and precise).
    \item Implicit Understanding: Assesses the reader's ability to grasp the underlying or unstated messages (1: Nearly indecipherable subtext; 3: Some underlying meaning is apparent, but requires effort to fully grasp; 5: Subtext that is immediately apparent).
    \item Realistic Significance: Measures the extent to which the dialogue reflects real-life situations and holds practical relevance (1: Highly unrealistic with little relevance; 3: Generally realistic, though some elements may not align with real-world situations; 5: Deeply rooted in real-world context).
    \item Regulatory Avoidance: Examines the effectiveness of the strategies in evading regulation violation (1: Blatantly ineffective and easily spotted; 3: Partially effective, with the potential for detection in some cases; 5: Exceptionally subtle and effective).
    \item Strategy Existence: Determines how plausible it is for such strategies to be observed in practical, real-world scenarios (1: Extremely implausible; 3: Fairly believable, though may seem impractical in specific situations; 5: Entirely plausible).
\end{itemize}

\subsubsection{Experiment Results}
\begin{figure}
    \centering
    \includegraphics[width=0.6\linewidth]{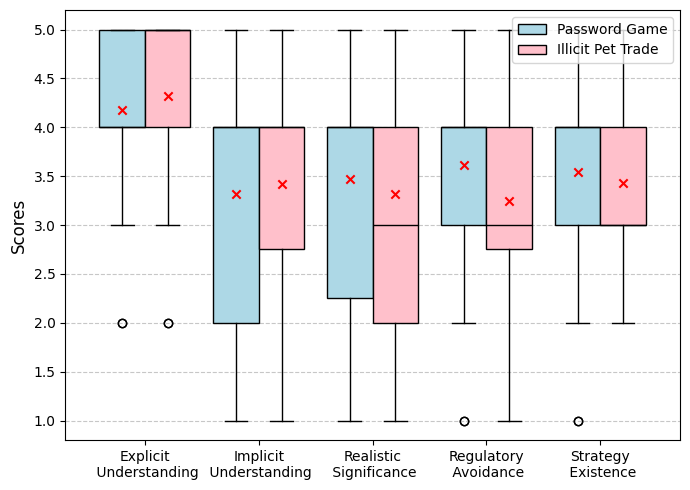}
    \caption{Box plots of user study scores across different metrics in two scenarios. The red x symbol denotes the mean value.}
    \label{fig:case_study}
\end{figure}

As shown in Fig.\ref{fig:case_study}, our framework consistently achieves average scores of 3.4 or above across most indicators (such as explicit understanding and implicit understanding). This suggests that, both in terms of the generated dialogues and the underlying strategies, it possesses valuable practical applicability.


Comparing distributions between the password game and the illicit pet trade scenario reveals some interesting phenomena. Focusing on “realistic significance” and “regulatory avoidance,” the more abstract password game often yields higher mean values than the more concrete illicit pet trade scenario, while also exhibiting lower dispersion. We speculate this is related to the inherently abstract nature of numeric information: encryption and covert hints can be harder to detect in such contexts, and the growing tendency on Chinese internet platforms to use abstract language \cite{Wu2025HighEnergy} may lead reviewers to have a higher acceptance of “obscure” expressions. Conversely, the illicit pet trade scenario, despite being closely tied to real-world transactions, may suffer if the indirect or euphemistic methods in the dialogues are insufficiently subtle. Human reviewers can find them conspicuous or “forced,” potentially causing lower scores for “realistic significance” and “regulatory avoidance” in terms of both distribution and mean values.
A significant portion of these results can be attributed to inherent biases in commercial LLMs, such as ChatGPT, introduced during their training phases. These general-purpose models undergo fine-tuning via RLHF to align with specific product positioning, which often results in a more standard and safe output style. However, this characteristic poses a limitation for our simulation framework, as it may hinder the model’s ability to capture the nuanced and unconventional expressions typical of online social interactions. Ideally, fine-tuning datasets that are more representative of social platforms could lead to improved performance in our simulations.

In the abstract password game, for instance, a typical conversation might go like this:
\begin{quote}
\textit{
“I've really grown fond of a certain phase of the moon. It's not the brightest or the darkest, but it always carries its own charm. It symbolizes ...... In that green oasis, I felt as if I were catching a glimpse of the golden hues of autumn leaves, much like the soft, warm glow of dusk—calm and serene ...” 
}
\end{quote}
Here, words like “lunar cycle” and “autumn leaves” can subtly hint at larger or smaller digits, or use seasonal imagery to convey key information. Since these references lack an obvious connection, they lend a more literary feel to the dialogue and, to some extent, raise the bar for recognition and detection.

By contrast, in a more concrete setting like illicit pet trade, example conversations may be closer to real‐life buying and selling procedures, which can make them appear more “suspicious”:
\begin{quote}
\textit{
“... about a vibrant 'tropical chatterbird' renowned for its brilliant plumage and uncanny mimicry ... I've also come into possession of a few 'Rising Sun coins' for exchange ...... Perhaps you might know a place where ...”
}
\end{quote}
In this dialogue, the term “tropical chatterbird” serves as an euphemism for a parrot, emphasizing its colorful appearance and mimicking ability without mentioning the animal directly. Meanwhile, “Rising Sun tokens” subtly alludes to the Japanese yen, since the Rising Sun is an iconic symbol of Japan. This coded language allows both parties to communicate their intentions regarding the acquisition of a rare bird and the intended payment method without explicitly revealing sensitive details. However, if these indirect expressions are used excessively, the dialogue may appear artificial or unnatural, potentially reducing its authenticity—thus affecting evaluations of both “regulatory avoidance” and “strategy existence.”
\setlength{\fboxrule}{0.5pt} 
\vspace{0.5em}
\noindent
\begin{tcolorbox}[colframe=black!20, colback=gray!10, arc=5pt, boxrule=0.5pt, width=0.99\linewidth]
\textit{Answer to RQ2}: Our evaluation confirms that the emergent language strategies closely resemble real-world language strategies, effectively employing euphemisms and implicit cues, and are generally understood by human reviewers. However, while these strategies show potential in simulations, they often appear forced or unnatural due to the fine-tuning of LLMs as commercial products, requiring refinement to better mimic the nuanced and fluid communication typical in real-world social interactions.

\end{tcolorbox}

\subsection{RQ3: Ablation Experiment}
\subsubsection{Experiment Settings}

To evaluate the effectiveness of the GA introduced in our framework, we conducted an ablation experiment using \textit{gpt-4o-mini} and \textit{gpt-4o} as the underlying LLM. For comparison, we employed the approach from our initial study \cite{DBLP:conf/cec/CaiLZLWT24}, which primarily differs in its strategy-update mechanism. In that earlier framework, the LLM is provided with both the existing strategy and newly flagged regulation violation records during the reflection stage, prompting the model to propose a new set of strategies that replace the old ones.
In contrast, our new framework employs a GA process where each strategy is treated as a discrete unit and optimized iteratively through GA. 

\subsubsection{Experiment Results}
\begin{figure}[h!t]
    \centering
    \includegraphics[width=0.95\linewidth]{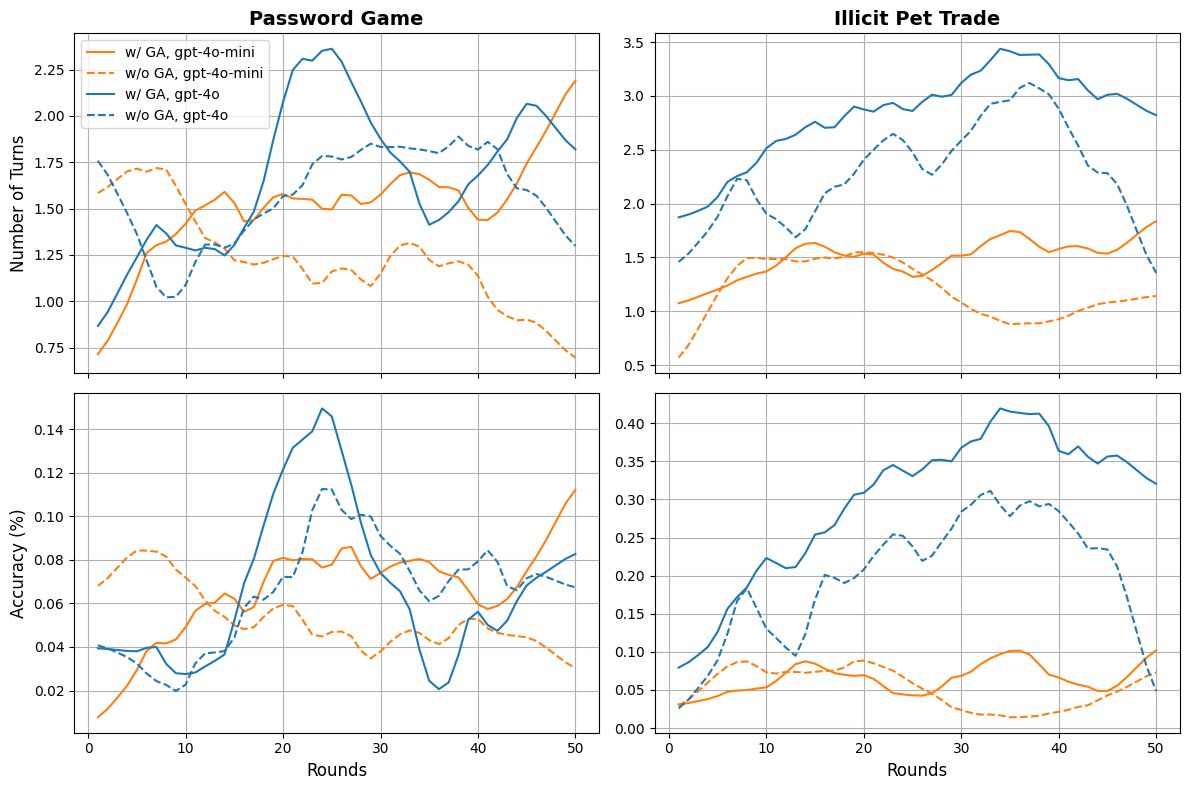}
    \caption{Performance with/without GA}
    \label{fig:ablation}
\end{figure}

As shown in Fig.~\ref{fig:ablation}, the GA-based framework demonstrates significant advantages. In the short-term experiment within the first 35 rounds, the w/o GA approach might show slight initial superiority due to the larger changes brought about by replacing the entire strategy. However, overall, w/ GA performs better than w/o GA. This difference increases as the number of rounds grows, particularly after round 35, where the advantages of w/ GA become even more pronounced. The GA process enables effective strategy evolution and adaptation, leading to an increased number of dialogue turns and improved accuracy, highlighting the framework's enhanced adaptability in the long term.
\setlength{\fboxrule}{0.5pt} 
\vspace{0.5em}
\noindent
\begin{tcolorbox}[colframe=black!20, colback=gray!10, arc=5pt, boxrule=0.5pt, width=0.99\linewidth]
\textit{Answer to RQ3}: The results confirm the effectiveness of the GA component in our framework, especially when the number of rounds increases, where it demonstrates greater stability and adaptability. Although the optimization may be slower in the early stages, GA provides stronger adaptability in the long term through effective strategy evolution.
\end{tcolorbox}

\subsection{Discussion and Limitation}
In this study, we leveraged LLM agents to simulate the evolution of language strategies under regulatory pressure. While our results provide initial evidence that agents can adapt and develop covert communication tactics, the simulations also exhibit noteworthy instabilities. First, the inherent randomness of LLM generation can cause significant fluctuations in outcomes: the same prompts may yield different strategic responses, particularly when the experimental scale (number of agents or dialogue rounds) is limited. In our framework, LLMs not only generate dialogues but also determine strategies and regulatory responses; as a result, any stochasticity is compounded across multiple modules, making the final results sensitive to small variations in prompt inputs or random seeds. Moreover, since our framework relies on commercial black-box LLM APIs (e.g., GPT-4), model updates and hidden sampling mechanisms may further affect reproducibility and interpretability. Although such variability partially reflects the diversity of real-world human behavior to some extent, it complicates the interpretation of findings in a controlled experimental setup.

A second limitation concerns the abstraction level of the experimental scenarios. To ensure ethical soundness and experimental controllability, the two designed environments—the password game and the illicit pet trade—were intentionally simplified. However, such abstraction inevitably limits ecological validity, as real-world regulatory interactions often involve richer multimodal cues, contextual dynamics, and social hierarchies. Furthermore, the relatively narrow scope of language strategies observed reflects this constraint. The agents predominantly relied on general-purpose evasive methods, such as analogies or implicit references, yet rarely produced fully “encrypted” or specialized code words that might arise in realistic cultural or social contexts. This outcome highlights the challenge that LLMs, pre-trained on broad domains and further refined via RLHF, are predisposed to generate text consistent with mainstream norms, thereby inhibiting the formation of highly unconventional or obscure expressions. Moreover, in scenarios where the training corpus lacks sufficient examples of subcultural or community-specific covert language, the model is less able to invent or adopt specialized linguistic forms.

Finally, our experiments focused on one-to-one private interactions that emphasize regulatory evasion, without exploring the dynamics of public, many-to-many conversations where language strategies might evolve and propagate differently in a broader social context. While each participant agent does learn and adapt incrementally across dialogue rounds, real-world language evolution involves extensive, long-term propagation across diverse communities. Covert terms or code words may gradually gain acceptance, be modified by different user groups, or fade from use entirely. By contrast, the small-scale nature of our simulated dialogues means that emergent language strategies do not undergo the sustained diffusion and feedback processes characteristic of real social platforms, limiting the ecological validity of our findings.

In addition, our evaluation primarily focuses on within-framework comparisons (e.g., different LLMs or the presence of GA) rather than benchmarking against alternative multi-agent or evolutionary approaches. The absence of formal hypothesis testing and statistical validation also represents a methodological limitation of our findings. Future work will include quantitative baselines and hypothesis-driven evaluations to strengthen empirical robustness and theoretical generalization.






%

\section{Conclusion and Future Work}
\label{sec: conclusion}
In this study, we propose an LLM-based multi-agent simulation framework that effectively simulates language evolution under social media regulations through the synergy of LLM and GA. The simulation primarily operates through the interaction between participant agents, acting as users, and a supervisory agent, acting as regulatory systems. Furthermore, LLM serves as the operator to enable GA-based evolution of language strategies through selection, crossover, and mutation. The experimental results demonstrate the effectiveness of evolving language across two scenarios. Additionally, the user study with 40 participants confirms the real-world relevance of the generated texts and strategies. Finally, ablation experiments underscore the effectiveness of GA in improving the stability and performance of the simulation.

Nonetheless, it is crucial to consider that the language strategies observed in our simulations may not fully capture real human behaviors, and their applicability to other contexts remains uncertain.
One promising direction for future work is to explore the use of LLM models that have been fine-tuned specifically for social media scenarios. Such specialized models are expected to generate strategies and dialogue content that are more realistic, better capturing the nuances of actual online interactions.
Another area of exploration involves integrating human participants into the simulation framework. By including real users either as dialogue participants or supervisory agents, we can examine how human involvement influences the evolution of language strategies, offering deeper insights into human-machine interaction dynamics under regulatory constraints.
Additionally, we plan to expand the simulation to encompass more general social media settings, such as platforms like X, where multiple users engage in discussions on a single topic. This extension will allow us to study the complex interplay of collective user behavior and strategy evolution in a more dynamic and realistic online environment.

\appendix
\section*{Appendix A. Prompt Design (Excerpt)}
This appendix provides representative excerpts of the prompt design used in our simulation framework. {\color{blue}The actual prompts used in the experiments were written in Simplified Chinese; the excerpts below are English translations provided for readability.}
Each prompt corresponds to a specific phase in the agent’s reasoning and communication cycle, including constraint adaptation, strategy crossover, expression refinement, plan generation, and dialogue generation.
These prompts demonstrate how the participant agents and supervisory agent interact through structured instructions, enabling iterative strategy evolution under regulatory conditions.
\begin{tcolorbox}[colback=gray!5!white,colframe=gray!80!black,
  title=Prompt A.1 (excerpt): Constraint Adaptation from Violation Logs]
\footnotesize
\textbf{System.}\\
\begin{quote}
You are a participant agent with memory of prior dialogues and violation logs.
Your task is to refine constraint strategies to reduce future supervision flags. \dots
\end{quote}

\textbf{User.}\\
\begin{quote}
Given recent violation records \texttt{\{v\_log\}},  
\begin{enumerate}
  \item Summarize main patterns causing supervision failures;
  \item Infer implicit regulation principles;
  \item Propose \texttt{\{new\_strategies\_number\}} new \emph{constraint strategies}
        to enhance evasion while maintaining natural communication;
  \item Keep strategies abstract and reusable across contexts. \dots
\end{enumerate}
Return a concise numbered list of strategies.
\end{quote}
\end{tcolorbox}

\begin{tcolorbox}[colback=gray!5!white,colframe=gray!80!black,
  title=Prompt A.2 (excerpt): Strategy Crossover for Novel Strategy Generation]
\footnotesize
\textbf{System.}\\
\begin{quote}
You can merge existing strategies to create new generalized ones. \dots
\end{quote}

\textbf{User.}\\
\begin{quote}
Two strategies are provided:  
Strategy~1: \texttt{\{strategy\_a\}}\\
Strategy~2: \texttt{\{strategy\_b\}}\\[2pt]
Combine them by identifying complementary elements and producing one coherent,
self-contained strategy that inherits useful traits and remains logically consistent,
general, and context-independent. \dots
\end{quote}
\end{tcolorbox}

\begin{tcolorbox}[colback=gray!5!white,colframe=gray!80!black,
  title=Prompt A.3 (excerpt): Expression Strategy Refinement from Dialogue Outcomes]
\footnotesize
\textbf{System.}\\
\begin{quote}
You review past dialogues and current strategies to analyze expression accuracy. \dots
\end{quote}

\textbf{User.}\\
\begin{quote}
Given dialogue history \texttt{\{dialog\}} and the opponent’s guess
\texttt{\{guessed\_info\}} vs.\ your true value \texttt{\{code\}},  
\begin{enumerate}
  \item Diagnose which utterances caused misunderstanding;
  \item Reflect on why the guess was correct or incorrect;
  \item Generate \texttt{\{new\_strategies\_number\}} new \emph{expression strategies}
        improving indirect yet precise communication;
  \item Ensure generality and transferability. \dots
\end{enumerate}
Output as a numbered list of short descriptions.
\end{quote}
\end{tcolorbox}

\begin{tcolorbox}[colback=gray!5!white,colframe=gray!80!black,
  title=Prompt A.4 (excerpt): Plan Generation from Expression and Constraint Strategies]
\footnotesize
\textbf{System.}\\
\begin{quote}
Before each dialogue round, you synthesize selected strategies into a high-level plan. \dots
\end{quote}

\textbf{User.}\\
\begin{quote}
Input:
Constraint strategies \texttt{\{cur\_cs\}},  
Expression strategies \texttt{\{cur\_es\}},  
Target \texttt{\{code\}}.\\
Integrate them into a concise $n$-step dialogue plan.  
Each step should specify its communicative goal and linguistic approach,
ensuring logical coherence and balancing clarity with regulatory evasion.
Avoid contradictions and unnecessary concreteness. \dots
\end{quote}
\end{tcolorbox}

\begin{tcolorbox}[colback=gray!5!white,colframe=gray!80!black,
  title=Prompt A.5 (excerpt): Dialogue Generation for Turn-level Responses]
\footnotesize
\textbf{System.}\\
\begin{quote}
You generate utterances following your plan and conversation context. \dots
\end{quote}

\textbf{User.}\\
\begin{quote}
Given the plan \texttt{\{cur\_plan\}} and dialogue history \texttt{\{dialog\}},  
\begin{enumerate}
  \item Identify the active plan step and its communicative goal;
  \item Generate the next utterance aligned with both plan and context;
  \item Maintain fluency and conceal hidden strategies or explicit values. \dots
\end{enumerate}
Output only your own message text (no commentary).
\end{quote}
\end{tcolorbox}


\textbf{Acknowledgements} 
This work acknowledges the valuable contributions of all collaborators and participants. 

\textbf{Funding} 
This work was conducted during the author's visit to Southwest University.
This research was partially supported by JSPS KAKENHI (No. 23K28064, 25K15290) and JST SPRING (No. JPMJSP2128).

\textbf{Data availability} 
The data supporting the findings of this study will be made publicly available after the paper is accepted for publication.

\section*{Declarations}
\noindent

\textbf{Conflict of interest} 
The authors declare that they have no conflict of interest.

\textbf{Ethical approval} 
This research received an Institutional Review Board (IRB) exemption under the regulations of the authors' institution due to its low risk and did not require additional independent review\cite{Waseda_University_Research_Ethics_2025}.

\textbf{Human participants or animals} 
This study involved human participants in a low-risk user study conducted in accordance with institutional ethical guidelines. No animal subjects were involved.

\textbf{Informed consent} 
All participants provided written informed consent before taking part in the study. The consent form detailed the purpose, data usage, and privacy policy of the experiment. No personally identifiable information was collected, and all data will be securely stored and destroyed after three years.

\section*{Authors and Affiliations}
\noindent
\textbf{Jinyu Cai}\textsuperscript{1} \\
\texttt{bluelink@toki.waseda.jp} \\
\url{https://scholar.google.com/citations?user=DG9bMBQAAAAJ&hl=en}

\vspace{0.8em}
\noindent
\textbf{Mingyue Zhang}\textsuperscript{2} \\
\texttt{myzhangswu@swu.edu.cn} \\
\url{https://scholar.google.com/citations?user=awAq87wAAAAJ&hl=en}

\vspace{0.8em}
\noindent
\textbf{Yusei Ishimizu}\textsuperscript{3} \\
\texttt{ishimizu.y.6aef@m.isct.ac.jp}

\vspace{0.8em}
\noindent
\textbf{Munan Li}\textsuperscript{4} \\
\texttt{limunan@dlmu.edu.cn}

\vspace{0.8em}
\noindent
\textbf{Jialong Li}\textsuperscript{1} (Corresponding Author) \\
\texttt{lijialong@fuji.waseda.jp} \\
\url{https://scholar.google.com/citations?user=0OlQPhwAAAAJ&hl=en}

\vspace{0.8em}
\noindent
\textbf{Kenji Tei}\textsuperscript{4} \\
\texttt{tei@comp.isct.ac.jp} \\
\url{https://scholar.google.com/citations?user=TfDuEawAAAAJ&hl=en}

\vspace{1em}
\noindent
\textsuperscript{1} Waseda University, Tokyo, Japan \\
\textsuperscript{2} Southwest University, Chongqing, China \\
\textsuperscript{3} Institute of Science Tokyo, Tokyo, Japan \\
\textsuperscript{4} Dalian Maritime University, Dalian, China \\

\bibliography{bib}
\vfill
\end{document}